\documentclass[aps,prl,a4paper,10pt,twocolumn,showpacs,floatfix,superscriptaddress,amsmath,amsfonts,amssymb,preprintnumbers]{revtex4-1}
\usepackage{float}

\usepackage{dcolumn,graphicx,color,booktabs,microtype,afterpage}

\graphicspath{{./}{figure/}}
\usepackage[charter,greekuppercase=italicized]{mathdesign}
\usepackage{sidecap}
\renewcommand{\tablename}{Table}
\makeatletter\renewcommand{\fnum@figure}[1]{\figurename~\thefigure.~}\makeatother
\makeatletter\renewcommand{\fnum@table}[1]{\tablename~\thetable.}\makeatother

\newcount\hh \newcount\mm
\hh=\time \divide\hh by 60
\mm=\hh \multiply\mm by 60 \mm=-\mm
\advance\mm by \time
\def\now{\number\hh:\ifnum\mm<10{}0\fi\number\mm}

\usepackage[colorlinks,plainpages=false,linkcolor=blue,urlcolor=blue,citecolor=blue,pdfpagemode=UseNone,pdfstartview=FitBH]{hyperref}

\newcommand{\tcr}[1]{\textcolor{black}{#1}}

\begin{document}

\makeatletter\renewcommand{\ps@plain}{%
\def\@evenhead{\hfill\itshape\rightmark}%
\def\@oddhead{\itshape\leftmark\hfill}%
\renewcommand{\@evenfoot}{\hfill\small{--~\thepage~--}\hfill}%
\renewcommand{\@oddfoot}{\hfill\small{--~\thepage~--}\hfill}%
}\makeatother\pagestyle{plain}


\preprint{\textit{Preprint: \today, \now}} 

\title{Superconductivity and topological aspects of the rock-salt carbides NbC and TaC}
%
\author{T.\ Shang}\email[Corresponding authors:\\]{tian.shang@psi.ch}
\affiliation{Laboratory for Multiscale Materials Experiments, Paul Scherrer Institut, Villigen CH-5232, Switzerland}
\author{J.\ Z.\ Zhao}\email[Corresponding authors:\\]{jzzhao@swust.edu.cn}
\affiliation{Co-Innovation Center for New Energetic Materials, Southwest University of Science and Technology, Mianyang, 621010, People's Republic of China} 
\affiliation{Research Laboratory for Quantum Materials, Singapore University of Technology and Design, Singapore 487372, Singapore} 
\author{D.~J.~Gawryluk}
\affiliation{Laboratory for Multiscale Materials Experiments, Paul Scherrer Institut, Villigen CH-5232, Switzerland}
\author{M.\ Shi}
\affiliation{Swiss Light Source, Paul Scherrer Institut, Villigen CH-5232, Switzerland}
\author{M.\ Medarde}
\affiliation{Laboratory for Multiscale Materials Experiments, Paul Scherrer Institut, Villigen CH-5232, Switzerland}
\author{E.\ Pomjakushina}
\affiliation{Laboratory for Multiscale Materials Experiments, Paul Scherrer Institut, Villigen CH-5232, Switzerland}
\author{T.\ Shiroka}
\affiliation{Laboratorium f\"ur Festk\"orperphysik, ETH Z\"urich, CH-8093 Z\"urich, Switzerland}
\affiliation{Paul Scherrer Institut, CH-5232 Villigen PSI, Switzerland}
\begin{abstract}
	
Superconducting materials with a nontrivial 
band structure are potential candidates for topological superconductivity. 
Here, by combining muon-spin rotation and relaxation ($\mu$SR) methods with 
theoretical calculations, we investigate the superconducting and topological 
properties of the rock-salt-type compounds NbC and TaC (with$T_c$ = 11.5 
and 10.3\,K, respectively). 
At a macroscopic level, the magnetization and heat-capacity measurements 
under applied magnetic field provide an upper critical field of 1.93 
and 0.65\,T for NbC and TaC, respectively. 
The low-temperature superfluid density, determined by transverse-field 
$\mu$SR and electronic specific-heat data, suggest a fully-gapped superconducting 
state in both NbC and TaC, with a zero-temperature gap $\Delta_0 = 1.90$ 
and 1.45\,meV, and a magnetic penetration depth $\lambda_0$ = 141 and 77\,nm, respectively. 
Band-structure calculations suggest that the density of states at the 
Fermi level is dominated by the Nb $4d$- (or Ta $5d$-) orbitals,
which are strongly hybridized with the C $p$-orbitals to produce large 
cylinder-like Fermi surfaces, similar to those of high-$T_c$ 
iron-based superconductors. 
Without considering the spin-orbit coupling (SOC) effect, the first 
Brillouin zone contains three closed node lines in the bulk band structure, 
protected by time-reversal and space-inversion symmetry. When 
considering SOC, its effects in the NbC case appear rather modest. 
Therefore, the node lines may be preserved in NbC,
hence proposing it as a potential topological superconductor.
\end{abstract}


\maketitle\enlargethispage{3pt}

\vspace{-5pt}
\section{\label{sec:Introduction}Introduction}\enlargethispage{8pt}
Recently, the binary TMA materials have been widely studied due to their exotic physical properties. Here, TM represents an early transition metal 
(e.g., Cr, Nb, Mo, Ta, or W), while A represents a carbon-, pnictogen-, or chalcogen group element (e.g., C, P, or As). 
Some TMA materials are believed to exhibit unconventional superconductivity (SC). 
For instance, originally CrAs and MnP exhibit antiferromagnetic and ferromagnetic long-range order below 264\,K and 290\,K, respectively~\cite{Cheng2017}. Under applied pressure, the magnetic order is suppressed 
and a dome-like superconducting phase appears near the magnetic quantum critical point, suggesting an  unconventional SC pairing in these materials~\cite{We2014,Kotegawa2015,Cheng2015,Cheng2017,Park2019}. Beyond superconductivity, TMAs are among 
the best candidate materials for studying topological phenomena. Weyl fermions, originally predicted in high-energy physics~\cite{Weyl1929}, were 
recently discovered as quasiparticles in Ta(As,P) and Nb(As,P) crystals via angle-resolved photoemission experiments~\cite{Xu2015a,Xu2015b,Lv2015,Xu2016,Souma2016}. Later on, three-component fermions were experimentally observed in MoP and WC~\cite{,Lv2017,Ma2018}. Now, all the above mentioned TMA materials are known as Weyl- 
or topological semimetals. 

The materials where superconductivity coexists with a nontrivial topological band structure may exhibit emergent phenomena, such as topological superconductivity and Majorana fermions~\cite{Qi2011,Kitaev2001}. 
By applying external pressure, the topological semimetal MoP becomes a superconductor, 
whose $T_c$ raises 
up to 4\,K (above 90\,GPa)~\cite{Chi2018}, thus representing a candidate topological superconductor.  In contrast to MoP, the orthorhombic WP is a superconductor already at ambient pressure 
(below 0.8\,K)~\cite{Liu2019}, but its topological nature is not yet known. 
In addition to MoP and WP, the transition-metal carbides (TMCs), too, can exhibit 
a large variety of band topologies in their different structural forms. The TMCs provide an exciting family of candidate materials 
for studying the rich physics of topological SC and Majorana bound states.
Generally, TMCs with a 1:1 metal-carbon stoichiometric ratio adopt two types of crystal structures. One is the so-called $\delta$-phase, a centrosymmetric cubic structure with space group $Fm\overline{3}m$ (No.\,225), also known as rock-salt phase. The other is the $\gamma$-phase, a noncentrosymmetric he\-xa\-go\-nal structure with space group $P\overline{6}m2$ (No.\,187). The above mentioned topological MoP and WC semimetals crystallize in the $\gamma$-phase. 
In case of MoC, although its SC has been reported in the 1970s~\cite{Willens1967}, its physical properties have been overlooked due to difficulties in synthesizing clean samples. Only recently, the carbon-defective $\delta$- and $\gamma$-phase of MoC$_{1-x}$ (with $T_c$ = 14.3 and 8.5\,K) could be synthesized under \tcr{high-temperature high-pressure conditions (1700\,$^\circ$C, 6--17\,GPa)}, and their crystal structures and physical properties be studied via different techniques~\cite{Yamaura2006,Sathish2012,Sathish2014}. Moreover, in both $\delta$- and $\gamma$-MoC, first-principle calculations could show the coexistence of SC with nontrivial band topology~\cite{Huang2018}. 

Other TMCs, such as $\delta$-type VC and CrC, are also predicted to be 
superconductors with nontrivial topological band structures and, therefore, 
are candidates for topological superconductivity~\cite{Zhan2019}. 
Although both VC and CrC should exhibit
relatively high critical temperatures~\cite{Isaev2007,Tutuncu2012,Kavitha2016,Szymanski2019}, 
to date experimental evidence is inconsistent. 
\tcr{For VC, the experimentally reported value ($T_c \sim 2$\,K) is an order of magnitude less 
	than the theoretically predicted one ($\sim$18\,K)~\cite{Ziegler1953,Szymanski2019}.
	As for CrC, 
	experimental evidence of SC is still missing.}
Besides MoC, also NbC and TaC show rock-salt type 
structures~\cite{Kazumas2008} and become superconductors below 11\,K~\cite{Willens1967,Williams1971,Toth1971}.
\tcr{Unlike MoP or MoC, where extremely high pressures are required to 
	induce SC, either during synthesis or as an external tuning parameter~\cite{Chi2018,Yamaura2006,Sathish2012,Sathish2014}, NbC and TaC are easy to 
	synthesize and show SC already at ambient pressure.}
However, to the best of our knowledge, 
none of the early works had a follow-up regarding 
the microscopic investigation of their superconducting properties. 
\tcr{Considering that $\delta$-type VC and CrC show nontrivial 
	topological band structures, we expect the isostructural NbC and TaC, 
	too, to exhibit similar features.} 

In this paper, we report on an extensive study of the superconducting properties of the NbC and TaC carbides, carried out via 
magnetization, heat-capacity, and muon-spin relaxation/rotation 
($\mu$SR) measurements. In addition, we also present numerical density-functional-theory (DFT) band-structure
calculations. We find that both NbC and TaC exhibit a fully-gapped superconducting state, while their electronic band structures suggest that, 
without considering spin-orbit coupling (SOC), both compounds are nodal line semimetals.  
After taking SOC into account, the degenerate bands are gapped out, 
except for six Dirac points on the high-symmetry lines.
Therefore, these two
carbides are potential candidates for studying topological SC and its associated Majorana bound states.

\section{Experimental and numerical methods\label{sec:details}}\enlargethispage{8pt}
The samples consisted of high-purity NbC and TaC (99+\%) powders 
acquired from ChemPUR. 
For the heat-capacity and $\mu$SR measurements the powders were pressed 
into small pellets, while for the magnetization measurements loose powders 
were used. 
Room-temperature x-ray powder diffraction (XRD) measurements were 
performed on a Bruker D8 diffractometer using Cu K$\alpha$ 
radiation. The magnetic susceptibility, 
and heat capacity measurements were performed on a 7-T Quantum Design magnetic 
property measurement system (MPMS-7) and a 9-T physical property measurement system (PPMS-9). 
The $\mu$SR measurements were carried out at the general-purpose-surface-muon (GPS) spectrometer 
of the Swiss muon source at Paul Scherrer Institut, Villigen, Switzerland~\cite{Amato2017}. 
The $\mu$SR data were analysed by means of the \texttt{musrfit} software package~\cite{Suter2012}.

The electronic band structures of NaC and TaC were calculated via the 
density functional theory, within the generalized gradient approximation 
(GGA) of Perdew-Burke-Ernzerhof (PBE) realization~\cite{Perdew:1996iq}, 
as implemented in the Vienna ab-initio Simulation Package (VASP)~\cite{Kresse:1996kl,Kresse:1996vk}. The projector augmented wave (PAW) pseudopotentials were adopted for the calculation~\cite{Kresse:1999wc,Blochl:1994zz}. Electrons belonging to the outer atomic configuration 
were treated as valence electrons, here corresponding to 5 electrons in Ta ($5d^36s^2$), 11 electrons in Nb ($4p^64d^45s$), and 4 electrons in C ($2s^22p^2$). The kinetic energy cutoff was fixed to 550\,eV. 
For the self-consistent calculation, the Brillouin zone integration was performed on a $\Gamma$-centered mesh of $20 \times 20 \times 20$ $k$-points. The spin-orbit coupling was taken 
into account by using a scalar relativistic approximation. 
In our calculations we used the lattice parameters and the atomic positions   experimentally determined from the Rietveld refinements. 

\section{\label{sec:results }Results and discussion}\enlargethispage{8pt} 
\subsection{\label{ssec:structure} Crystal structure}

\begin{figure}[!bht]
	\centering 
	\includegraphics[width=0.48\textwidth,angle=0]{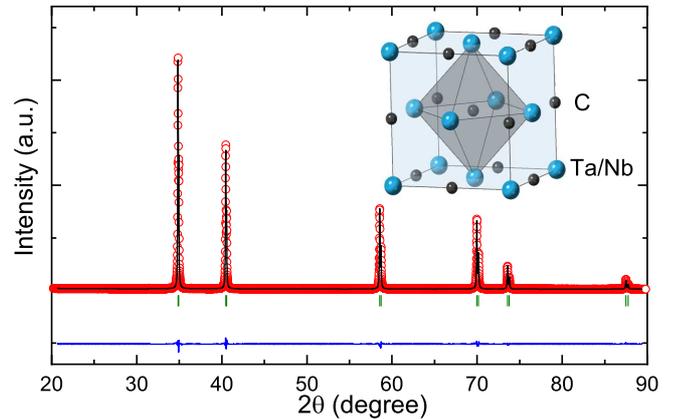} 
	\caption{\label{fig:XRD} Room-temperature x-ray powder diffraction 
		pattern and Rietveld refinements for TaC. The open red circles 
		and the solid black line represent the experimental pattern and the 
		refinement profile, respectively. The blue line at the bottom
		shows the residuals, i.e., the difference between the calculated 
		and the experimental data. The vertical bars mark the calculated Bragg-peak 
		positions. The cubic crystal structure (unit cell) is shown in the inset.}
\end{figure}
%

\begin{figure}[th]
	\centering
	\includegraphics[width=0.48\textwidth,angle=0]{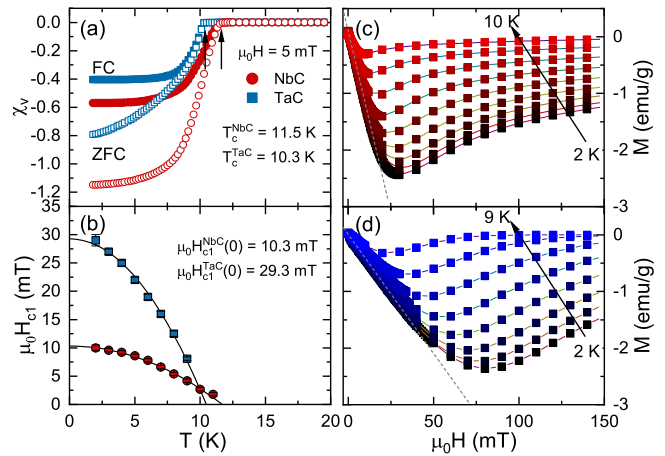}
	\caption{\label{fig:Chi}(a) Magnetic susceptibility of NbC and TaC 
		vs.\ temperature, measured in an applied field of 5\,mT using both 
		ZFC and FC protocols.
		(b) Estimated lower critical field $\mu_{0}H_{c1}$ vs.\ temperature 
		for NbC and TaC. The solid lines are fits to $\mu_{0}H_{c1}(T) =\mu_{0}H_{c1}(0)[1-(T/T_{c})^2]$. 
		\tcr{Field-dependent magnetization recorded 
			at various temperatures up to $T_c$ are shown in (c) for NbC and (d) for TaC}. For each temperature, the lower critical 
		field $\mu_{0}H_{c1}$ was determined as the value where $M(H)$ starts 
		deviating from linearity (see dashed lines).}
\end{figure}
%

The \tcr{phase} purity and the crystal structure of NbC and TaC powders were 
checked via XRD at room temperature. Figure~\ref{fig:XRD} shows the 
XRD pattern of TaC (with NbC showing a similar pattern), analyzed by 
means of the FullProf Rietveld-analysis suite~\cite{Carvajal1993}. 
Consistent with previous neutron scattering results~\cite{Kazumas2008}, 
we find that NbC and TaC powders crystallize in the simple face-centered-cubic (FCC) NaCl-type structure, with space group
$Fm\overline{3}m$ (No.\ 225) (see crystal structure in the inset of 
Fig.~\ref{fig:XRD}). There are only two atomic positions in the unit 
cell: $4a$ (0, 0, 0) for the Nb/Ta atoms, 
and $4b$ (0.5, 0.5, 0.5) for the C atoms.
The refined lattice parameters, $a = 4.4676(1)$\,\AA\ (NbC) and 
$a = 4.4557(1)$\,\AA\ (TaC), are consistent with the values determined 
from neutron scattering~\cite{Kazumas2008}. No impurity phases could be 
detected in either case, thus indicating a good sample quality.

\subsection{\label{ssec:sus}Magnetization measurements}
%
\begin{figure*}[htp]
	\centering
	\includegraphics[width=0.9\textwidth,angle= 0]{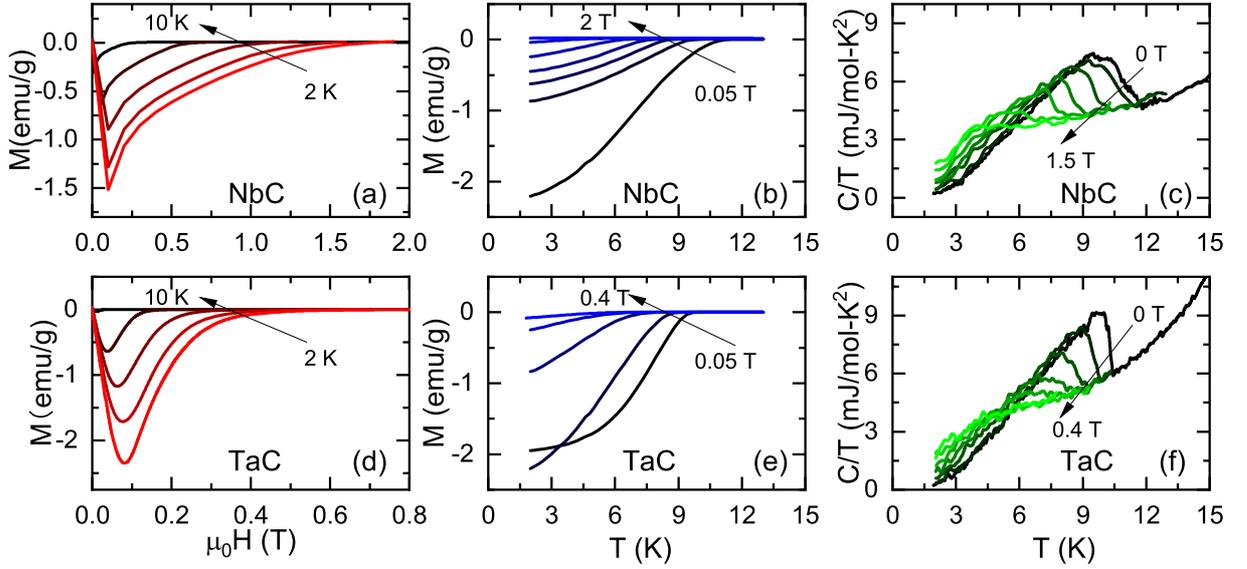}
	\caption{\label{fig:Hc2_determ}(a) Field-dependent magnetization $M(H,T)$ collected at various temperatures,
		(b) temperature-dependent magnetization $M(T,H)$, and (c) specific heat $C(T,H)/T$ measured in 
		various applied magnetic fields for NbC. The analogous 
		results for the TaC samples are presented in the panels (d)-(f), respectively. }
\end{figure*}
%

The superconductivity of the NbC and TaC powders was first characterized 
by magnetic susceptibility measurements, carried out in a 5-mT field, 
using both field-cooled (FC) and zero-field-cooled (ZFC) protocols.
As indicated by the arrows in Fig.~\ref{fig:Chi}(a), a clear 
diamagnetic signal appears below the superconducting 
transition at $T_c$ = 11.5\,K and 10.3\,K for NbC and TaC, respectively. 
The $M(H)$ data for NbC and TaC are plotted in Fig.~\ref{fig:Chi}(c) and Fig.~\ref{fig:Chi}(d), respectively. 
In both cases, the field-dependent magnetization $M(H)$, collected at various temperatures up to $T_c$, allowed us to determine the lower 
critical field $\mu_{0}H_{c1}$.
The estimated $\mu_{0} H_{c1}$ values as a function 
of temperature are summarized in Fig.~\ref{fig:Chi}(b).
The solid lines represent fits to $\mu_{0}H_{c1}(T) = \mu_{0}H_{c1}(0)[1-(T/T_{c})^2]$ and yield a lower critical field 
of 10.3(3) and 29.3(3)\,mT for NbC and TaC, respectively.  

\subsection{\label{ssec:critical_field} Upper critical fields}  

\begin{figure}[htp]
	\centering
	\includegraphics[width=0.49\textwidth,angle= 0]{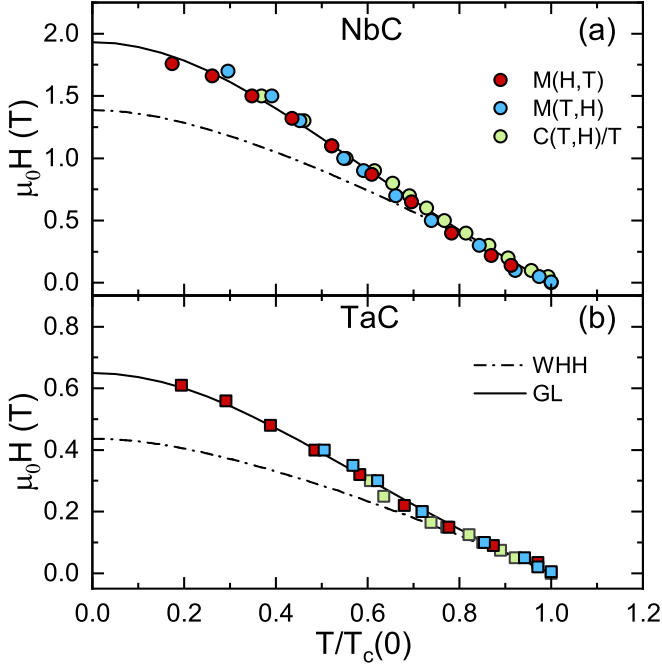}
	\caption{\label{fig:Hc2}Summary of the upper critical field data $\mu_{0}H_{c2}$ vs.\ reduced temperature $T_c/T_c(0)$, as determined from field-dependent magnetization $M(H,T)$, temperature-dependent magnetization $M(T,H)$, and specific heat $C(T,H)/T$ for (a) NbC  and (b) TaC, respectively. 
		For $M(H,T)$, the $H_{c2}$ value was determined as the field where 
		the diamagnetic signal is suppressed. Two different models, including 
		an effective GL- (solid lines) and a WHH model (dash-dotted lines), 
		were used to analyze the $\mu_{0}H_{c2}(T)$ data.} 
\end{figure}

%
\begin{figure}[!htp]
	\centering
	\includegraphics[width=0.46\textwidth,angle=0]{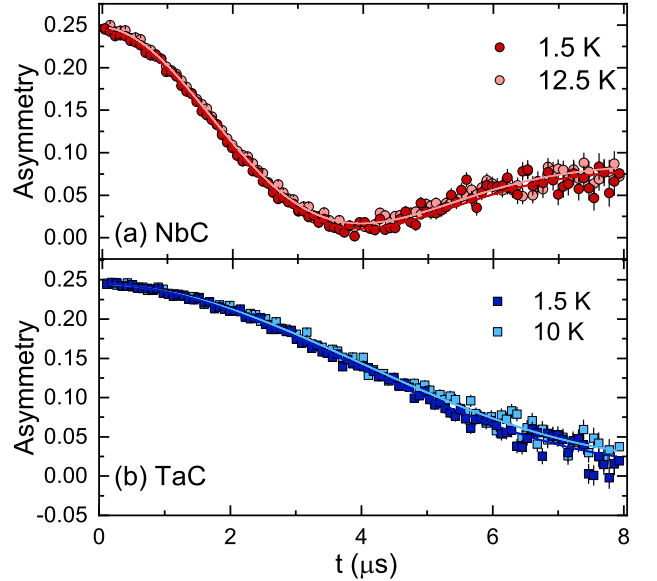}
	\vspace{-2ex}%
	\caption{\label{fig:ZF_muSR}ZF-$\mu$SR spectra of (a) NbC and (b) 
		TaC, recorded in the su\-per\-con\-duc\-ting- and the normal states. 
		Solid lines are fits using the equation described in the text. 
		None of the datasets shows evident
		changes with temperature.}
\end{figure}

To estimate the upper critical field $\mu_0 H_{c2}$ of NbC and TaC, 
temperature-dependent magnetization $M(T,H)$ and specific-heat 
$C(T,H)$/$T$ measurements, at various applied magnetic fields, 
as well as field-dependent magnetization $M(H,T)$ measurements, at 
various temperatures, were performed.
As shown in Figs.~\ref{fig:Hc2_determ}(a) and (d), in both samples, 
the diamagnetic signal progressively disappears as the applied 
magnetic field exceeds the upper critical field. In Figs.~\ref{fig:Hc2_determ}(b)-(c) and in Figs.~\ref{fig:Hc2_determ}(e)-(f), the superconducting transition in both $M(T)$ and $C(T)$/$T$ datasets shifts 
towards lower temperatures as the applied field increases.
In zero magnetic field, $T_c$ = 10.8 and 10.2\,K, determined from $C(T)$/$T$ for NbC and TaC, are consistent with the $T_c$ values determined from magnetic susceptibility data 
[Fig.~\ref{fig:Chi}(a)].
The upper critical fields determined from $M(H,T)$, $C(T,H)/T$, and 
$M(T,H)$ data are summarized in Figs.~\ref{fig:Hc2}(a) and (b) as a 
function of the reduced superconducting transition temperature $T_c$/$T_c$(0) 
for NbC and TaC, respectively. The $\mu_0 H_{c2}(T)$ behavior was 
analyzed by means of the Werthamer-Helfand-Hohenberg (WHH) and 
Ginzburg-Landau (GL) models~\cite{Werthamer1966,Zhu2008,Tinkham1996}. 
Both models describe the experimental data very well 
at low fields. However, at higher applied fields, the fits to the 
WHH model (dash-dotted lines) deviate significantly from the 
data, clearly providing underestimated $\mu_0 H_{c2}$ values.
Conversely, the GL model (solid lines) shows a remarkable 
agreement with the experimental data also at higher fields.
Fits to the GL model, which reproduces the data very well across the 
full temperature range, provide $\mu_0 H_{c2}^\mathrm{GL}(0)$ = 1.93(1)\,T 
and 0.65(1)\,T for NbC and TaC, respectively.



\subsection{\label{ssec:ZF_muSR}Zero-field \texorpdfstring{$\mu$SR}{MuSR}}
Zero-field (ZF) $\mu$SR measurements are quite suited for detecting 
possible magnetic order or magnetic fluctuations, as well as time-reversal 
symmetry breaking in the superconducting state.
Here, we performed ZF-$\mu$SR measurements in both the normal- and 
the superconducting states of NbC and TaC, with 
representative 
spectra collected above and below $T_c$ being shown in Fig.~\ref{fig:ZF_muSR}.
In either case, neither coherent oscillations nor fast decays could 
be identified, hence implying the lack of any magnetic order or fluctuations. 
The weak muon-spin relaxation we detect in absence of an external magnetic 
field is mainly due to the randomly oriented nuclear moments, 
which can be modeled by a Gaussian Kubo-Toyabe relaxation function $G_\mathrm{KT} = [\frac{1}{3} + \frac{2}{3}(1 -\sigma_\mathrm{ZF}^{2}t^{2})\,\mathrm{e}^{-\sigma_\mathrm{ZF}^{2}t^{2}/2}]$
~\cite{Kubo1967,Yaouanc2011}.
Here, $\sigma_\mathrm{ZF}$ is the zero-field Gaussian relaxation rate. 
%
\begin{figure}[!htp]
	\centering
	\includegraphics[width=0.48\textwidth,angle= 0]{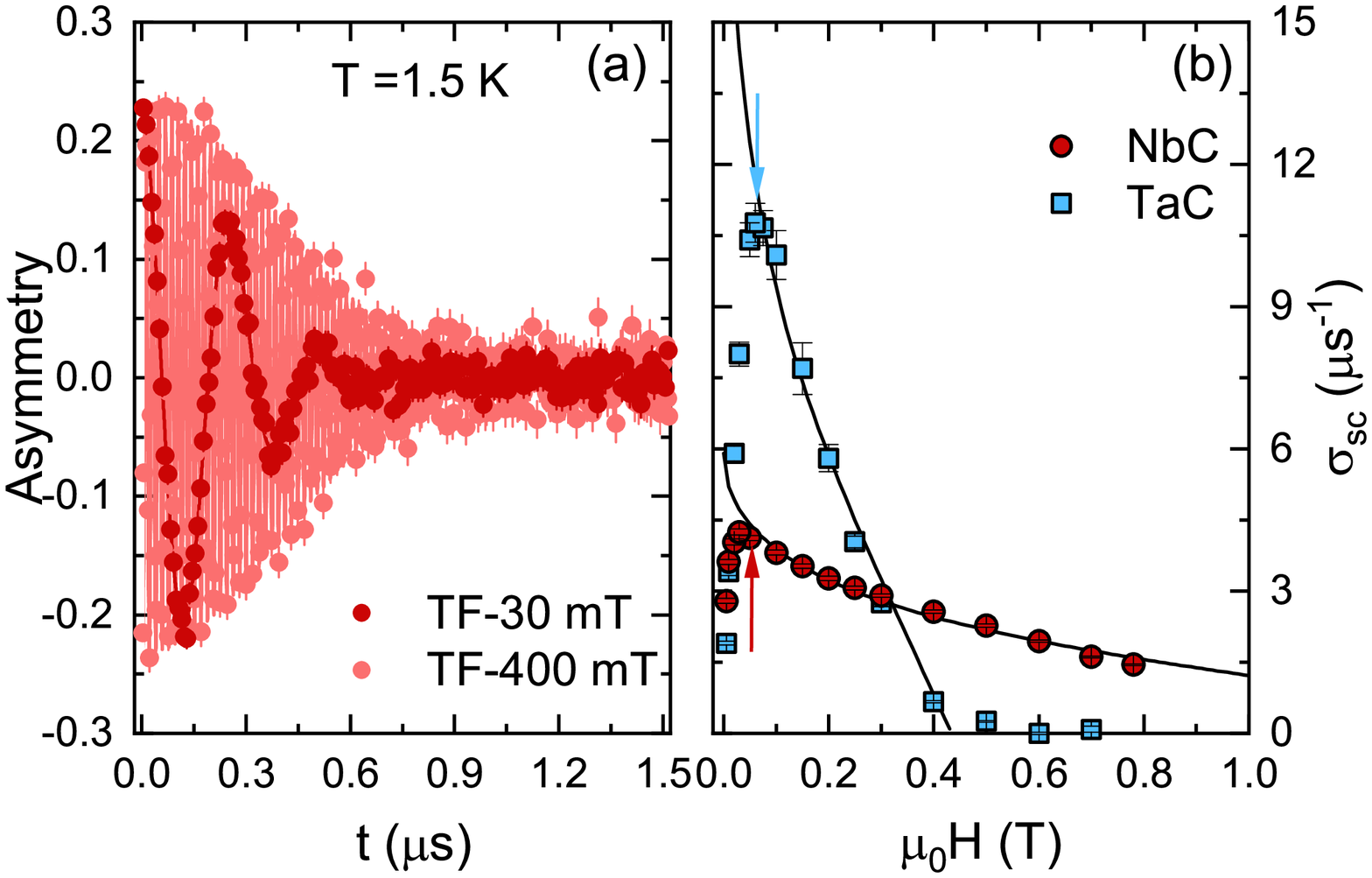}
	\caption{\label{fig:TF-muSR_H}(a) TF-$\mu$SR time spectra collected 
		in the superconducting state of NbC (at $T = 1.5$\,K) in an applied 
		field of 30 and 400\,mT. Similar results were obtained for TaC.  
		(b) Field-dependent Gaussian relaxation rate $\sigma_\mathrm{sc}(H)$ 
		for NbC and TaC. Solid lines are fits to Eq.~\eqref{eq:TF_muSR_H}, as 
		described in the text. The arrows indicate the field values used in the temperature-dependent TF-$\mu$SR studies of NbC and TaC, 50 and 60\,mT respectively.}
\end{figure}
%
%
The solid lines in Fig.~\ref{fig:ZF_muSR} represent fits to the data by considering also an additional zero-field Lorentzian relaxation $\Lambda_\mathrm{ZF}$, i.e., $A_\mathrm{ZF}(t) = A_\mathrm{s} G_\mathrm{KT} \mathrm{e}^{-\Lambda_\mathrm{ZF} t} + A_\mathrm{bg}$. 
Here $A_\mathrm{s}$ and $A_\mathrm{bg}$ represent the initial muon-spin asymmetries 
for muons implanted in the sample and sample holder (\tcr{copper}), respectively. 
In either carbide compound, the relaxations in both the normal- and the 
superconducting states are almost identical, as demonstrated by the practically overlapping ZF-$\mu$SR spectra above and below $T_c$. 
This lack of evidence for an additional $\mu$SR relaxation below $T_c$ 
excludes a possible time-reversal symmetry breaking in the superconducting 
state of NbC or TaC \tcr{and is supported also by the theoretical 
	arguments discussed in Sec.~\ref{ssec:DFT}}.  
The larger Gaussian relaxation rate of NbC ($\sigma_\mathrm{ZF} = 0.432$\,$\mu$s$^{-1}$) compared to TaC ($\sigma_\mathrm{ZF}$ = 0.183\,$\mu$s$^{-1}$) reflects the  larger nuclear magnetic moment of $^{93}$Nb (6.2\,$\mu_n$) compared to $^{181}$Ta (2.4\,$\mu_n$), the two ratios (in presence of similar structures) being almost the same, $\sim 2.5$. Finally, both samples show a very small 
Lorentzian relaxation ($\Lambda_\mathrm{ZF}$ $\sim$ 0.011\,$\mu$s$^{-1}$).
\tcr{%
	Unlike in pure Nb, where there is a clear muon hopping (with a mobility 
	minimum at 50\,K) and where a dynamic Kubo-Toyabe function describes 
	the ZF-$\mu$SR spectra quite well~\cite{Grassellino2013}, in the NbC 
	and TaC case, no dynamic features were observed.}


\subsection{\label{ssec:TF_muSR}Transverse-field \texorpdfstring{$\mu$SR}{MuSR}}
To investigate the superconducting properties of NbC and TaC at a microscopic 
level, we carried out systematic trans\-verse\--field (TF) $\mu$SR measurements. 
Generally, performing such measurements on type-II superconductors 
requires 
an applied magnetic field which exceeds 
$\mu_{0}H_{c1}$, thus allowing one to quantify the additional 
field-distribution broadening due to the flux-line lattice (FLL). 
Ideally, the optimal field value for such measurements is determined 
experimentally, via field-dependent $\mu$SR depolarization-rate 
measurements in the superconducting state. 
To track the additional field-distribution broadening due 
to the FLL in the mixed superconducting state, a magnetic field (up to 780\,mT) was applied in the normal state and then 
the sample was cooled down to 1.5\,K, where the TF-$\mu$SR spectra were collected. 

As an example, Fig.~\ref{fig:TF-muSR_H}(a) shows the spectra collected 
in an applied field of 30 and 400\,mT in NbC, with TaC showing similar 
features. The solid lines are fits using the same model as described 
in Eq.~\eqref{eq:TF_muSR_H} below.
%
%
\begin{figure}[!thp]
	\centering
	\includegraphics[width=0.48\textwidth,angle= 0]{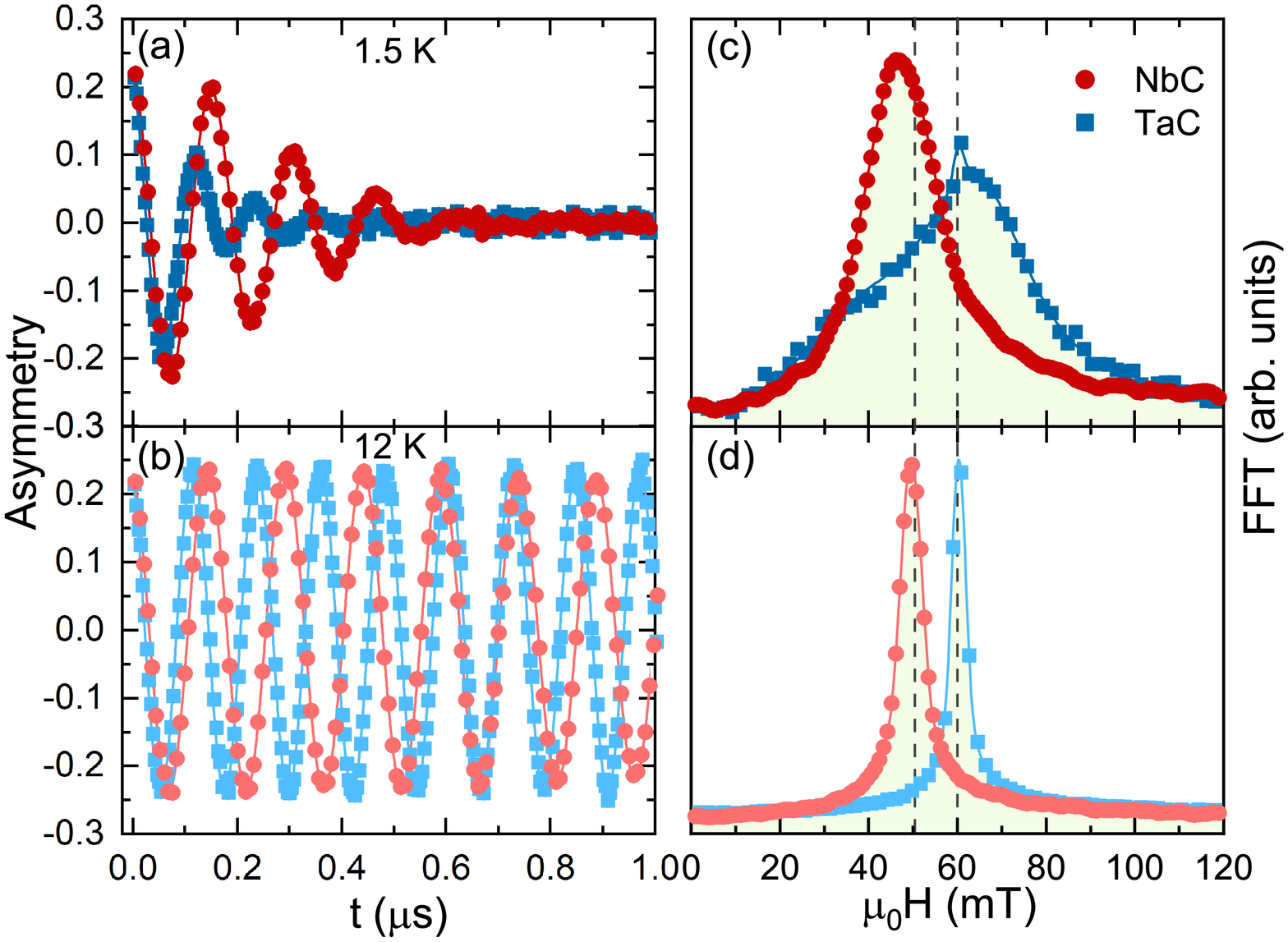}
	\caption{\label{fig:TF-muSR_T}TF-$\mu$SR spectra collected in the (a) superconducting- (1.5\,K) and (b) normal-state (12\,K), 
		in an applied field of 50\,mT for NbC and 60\,mT for TaC. 
		Fast Fourier transforms of the relevant time spectra at (c) 1.5\,K 
		and (d) 12\,K. The solid lines through the data are fits to 
		Eq.~(\ref{eq:TF_muSR}); the vertical dashed lines mark the applied 
		magnetic fields. Note the clear diamagnetic shift and the field 
		broadening in (c), typical of the superconducting phase.}
\end{figure}
%
The resulting superconducting Gaussian relaxation rates $\sigma_\mathrm{sc}(H)$ 
are summarized in Fig.~\ref{fig:TF-muSR_H}(b). For both NbC and TaC, the relaxation rate decreases continuously when the applied field is 
larger than the lower critical field $\mu_{0}H_{c1}$.
Therefore, as indicated in Fig.~\ref{fig:TF-muSR_H}(b), fields 
of 50 and 60\,mT were chosen as suitable for the temperature-dependent 
TF-$\mu$SR studies of NbC and TaC, respectively. 
The field-dependent Gaussian relaxation rate $\sigma_\mathrm{sc}(H)$ 
can be described by the expression~\cite{Barford1988,Brandt2003}:
\begin{equation}
\label{eq:TF_muSR_H}
\sigma_\mathrm{sc} = 0.172 \frac{\gamma_{\mu} \Phi_0}{2\pi}(1-h)[1+1.21(1-\sqrt{h})^3]\lambda^{-2}, 
\end{equation} 
where $\lambda$ is the magnetic penetration depth, 
$\gamma_{\mu} = 2\pi \times 135.53$\,MHz/T is the muon 
gyromagnetic ratio, and $h = H_\mathrm{appl}/H_\mathrm{c2}$, 
with $H_\mathrm{appl}$ the 
applied magnetic field. The above expression is valid for type-II 
superconductors with $\kappa \ge 5$ in the $0.25/\kappa^{1.3} \lesssim h \le$ 1 
field range. With $\kappa \sim$ 13 and 5, and $h$ = 0.026 and 0.092 for NbC and TaC, both samples fulfill the above conditions. 
The solid lines in Fig.~\ref{fig:TF-muSR_H}(b) are fits to the above equation. The derived $\mu_0$$H_{c2}$ = 1.81(2) and 0.43(1)\,T and magnetic 
penetration depths $\lambda_0$ = 134(2) and 67(1)\,nm for NbC and TaC 
are comparable with the measured upper critical field values 
(see Fig.~\ref{fig:Hc2}) and those determined via 
temperature-dependent TF-$\mu$SR (see Fig.~\ref{fig:lambda}). 

TF-$\mu$SR spectra were also collected at various temperatures up 
to $T_c$ in a fixed applied field. Figures~\ref{fig:TF-muSR_T}(a) and (b) show representative TF-$\mu$SR spectra, collected  
below (1.5\,K) and above $T_c$ (12\,K), for both NbC and TaC. 
The additional field distribution broadening due to FLL in the mixed state is clearly reflected in the enhanced 
muon-spin depolarization below $T_c$. 
To describe the field distribution, the asymmetry of TF-$\mu$SR spectra can be modelled using: 
\begin{equation}
\label{eq:TF_muSR}
A_\mathrm{TF}(t) = \sum\limits_{i=1}^n A_i \cos(\gamma_{\mu} B_i t + \phi) e^{- \sigma_i^2 t^2/2} +
A_\mathrm{bg} \cos(\gamma_{\mu} B_\mathrm{bg} t + \phi).
\end{equation}
%
%
%
\begin{figure}[!thp]
	\centering
	\includegraphics[width=0.45\textwidth,angle= 0]{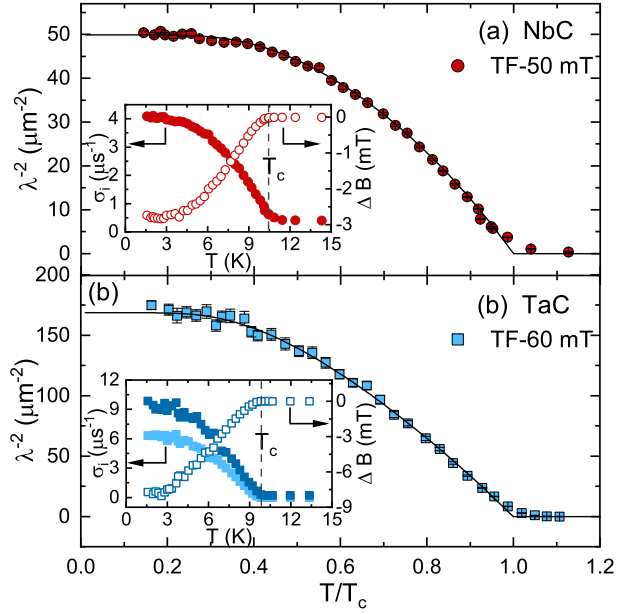}
	\caption{\label{fig:lambda} Superfluid density vs.\ temperature, as 
		determined from TF-$\mu$SR measurements in an applied magnetic field 
		of 50\,mT for (a) NbC and 60\,mT for (b) TaC. The insets show the 
		temperature dependence of the muon-spin relaxation rate $\sigma(T)$ 
		and of the diamagnetic shift 
		[$\Delta B (T) = \langle B \rangle - B_\mathrm{appl.}$]. 
		Two $\sigma$ values are required to describe the TF-$\mu$SR data   
		of TaC, while NbC requires only one $\sigma$ [see details in 
		Fig.~\ref{fig:TF-muSR_T}(c)].}
\end{figure}
%
%
Here $A_i$ and $A_\mathrm{bg}$ are the same as in ZF-$\mu$SR, with 
the latter term not undergoing any depolarization. $B_i$ and $B_\mathrm{bg}$ 
are the local fields sensed by implanted muons in the sample and sample 
holder, $\phi$ is a shared initial phase, and $\sigma_i$ 
is the Gaussian relaxation rate of the $i$th component. 

Figures~\ref{fig:TF-muSR_T}(c) and (d) show the fast Fourier transform 
(FFT) of the TF-$\mu$SR spectra
in Fig.~\ref{fig:TF-muSR_T}(a) and (b), respectively. As can be seen in Fig.~\ref{fig:TF-muSR_T}(c), the 
field distribution in TaC is much broader and asymmetric than in NbC, consistent with a larger muon-spin depolarization in TaC in 
Fig.~\ref{fig:TF-muSR_T}(a). 
Solid lines represent fits to Eq.~\eqref{eq:TF_muSR} using a single 
oscillation (i.e., $n = 1$) or two oscillations (i.e., $n = 2$) 
for NbC and TaC, respectively. 
The derived Gaussian relaxation rates as a function of temperature 
are summarized in the insets of Fig.~\ref{fig:lambda}, together with 
the diamagnetic shifts. 
Above $T_c$ the relaxation rate is small and temperature-independent, but 
below $T_c$ it starts to increase due to the formation of the FLL and the 
increase in superfluid density. In addition, a diamagnetic field shift appears in both samples below $T_c$.  

In case of multi-component oscillations, the first term in 
Eq.~\eqref{eq:TF_muSR} describes the field distribution as the sum of 
$n$ Gaussian relaxations (here $n = 2$ for TaC)~\cite{Maisuradze2009}:
\begin{equation}
\label{eq:TF_muSR_2}
P(B) = \gamma_{\mu} \sum\limits_{i=1}^2 \frac{A_i}{\sigma_i} \mathrm{exp}\left[-\frac{\gamma_{\mu}^2(B-B_i)^2}{2\sigma_i^2}\right].
\end{equation}
The first- and the second moments of the field distribution can be calculated by~\cite{Maisuradze2009}: 
\begin{equation}
\begin{aligned}
\label{eq:1st_moment}
&	\langle B \rangle =  \sum\limits_{i=1}^2 \frac{A_i B_i}{A_\mathrm{tot}},\quad \mathrm{and} \\
& \langle B^2 \rangle = \frac{\sigma_\mathrm{eff}^2}{\gamma_\mu^2} = \sum\limits_{i=1}^2 \frac{A_i}{A_\mathrm{tot}}\left[\frac{\sigma_i^2}{\gamma_{\mu}^2} + \left(B_i - \langle B \rangle\right)^2\right],
\end{aligned}
\end{equation}
%
%
\begin{figure}[th]
	\centering
	\includegraphics[width=0.43\textwidth,angle=0]{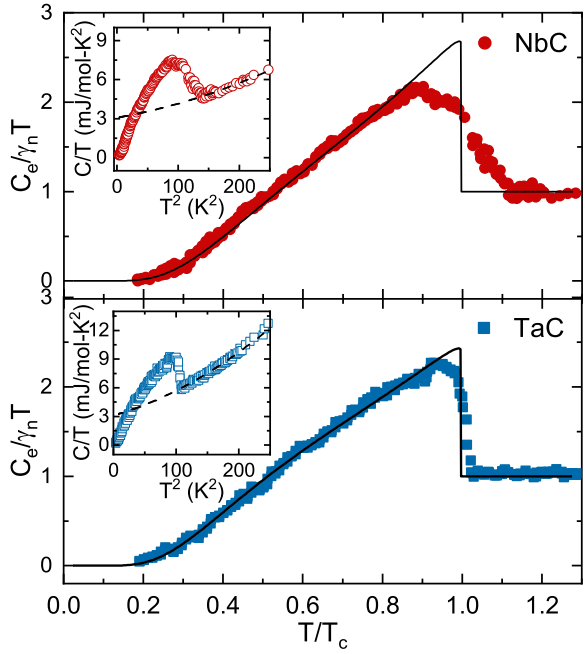}
	\vspace{-2ex}%
	\caption{\label{fig:Cp}%
		Normalized electronic specific heat $C_\mathrm{e}/\gamma_n T$ versus reduced temperature $T/T_c$ for NbC (a) and TaC (b). Here 
		$\gamma_n$ is the normal-state electronic specific-heat coefficient. The measured specific heat $C/T$ versus $T^2$ are shown in the insets. The 
		dashed-lines in the insets are fits to $C/T = \gamma_n + \beta T^2 + \delta T^4$ for $T > T_c$, while the solid lines in the main panel represents the  
		electronic specific heat calculated by considering a fully-gapped $s$-wave model for $T \le T_c$.}
\end{figure}
%
%
where $A_\mathrm{tot} = A_1 + A_2$.  After subtracting the nuclear relaxation rate $\sigma_\mathrm{n}$, according to $\sigma_\mathrm{sc} = \sqrt{\sigma_\mathrm{eff}^{2} - \sigma^{2}_\mathrm{n}}$, the superconducting Gaussian relaxation rate $\sigma_\mathrm{sc}$ can be 
extracted. The $\sigma_\mathrm{n}$ is considered to be temperature 
independent, as confirmed also by the ZF-$\mu$SR data in Fig.~\ref{fig:ZF_muSR}.  

For both NbC and TaC, the inverse-square of the magnetic penetration 
depth $\lambda^{-2}(T)$ [proportional to the superfluid density 
$\rho_\mathrm{sc}(T)$] is calculated by using Eq.~\eqref{eq:TF_muSR_H} 
and shown in the main panels of Fig.~\ref{fig:lambda} vs.\ the reduced $T/T_c$. 
Below $T_c/3$, the superfluid density is almost constant, 
thus excluding the possibility of superconducting gap nodes and 
indicating a fully-gapped SC in NbC and TaC. 
To gain more quantitative insights into the superconductivity of NbC 
and TaC, the superfluid density $\rho_\mathrm{sc}(T)$ was further 
analyzed by means of a fully-gapped $s$-wave model:
\begin{equation}
\label{eq:rhos}
\rho_\mathrm{sc}(T) =  1 + 2\int^{\infty}_{\Delta(T)} \frac{E}{\sqrt{E^2-\Delta^2(T)}} \frac{\partial f}{\partial E} dE.
\end{equation}
Here $f = (1+e^{E/k_\mathrm{B}T})^{-1}$ and $\Delta(T)$ are the Fermi- 
and the su\-per\-con\-duc\-ting\--gap functions. 
The $\Delta(T)$ is assumed to follow  $\Delta(T) = \Delta_0 \mathrm{tanh} \{ 1.82[1.018(T_\mathrm{c}/T-1)]^{0.51} \}$ 
\cite{Carrington2003}, where $\Delta_0$ is the zero-temperature superconducting gap value. 
The solid lines in the main panels of Fig.~\ref{fig:lambda} are fits to 
the above model with a single gap, which yield a zero-temperature  
gap values $\Delta_0$ = 1.90(2) and 1.45(1)\,meV, and magnetic penetration depths $\lambda_0$ = 141(2) and 77(1)\,nm for NbC and TaC, respectively.
Since in NbC, the $H_\mathrm{appl}/H_\mathrm{c2}$ $\ll$ 1, the magnetic 
penetration depth can also be calculated using 
$\sigma_\mathrm{sc}^2(T)/\gamma^2_{\mu} = 0.00371\, \Phi_0^2/\lambda^4(T)$~\cite{Brandt2003}, which gives a comparable $\lambda_0$ = 162(2)\,nm.

\tcr{
	According to the GL theory of superconductivity, the coherence length 
	$\xi$ can be calculated by using $\xi$ =  $\sqrt{\Phi_0/2\pi\,H_{c2}}$, 
	where $\Phi_0 = 2.07 \times 10^{3}$\,T~nm$^{2}$
	is the quantum of magnetic flux. With a bulk $\mu_{0}H_{c2}(0)$ = 1.93(1)\,T and 0.65(1)\,T, 
	the calculated $\xi(0)$ are 13.1(1)\,nm and 22.5(1)\,nm for NbC and TaC, respectively. 
	A GL parameter $\kappa = \lambda/\xi \sim$ 11 (NbC) and 3.4 (TaC), 
	much larger than the $1/\sqrt{2}$ threshold value, clearly confirms 
	type-II superconductivity in both NbC and TaC, consistent with the 
	magnetization results in Sec.~\ref{ssec:sus}.
}

\begin{figure}[ht]
	\centering
	\includegraphics[width = 0.45\textwidth]{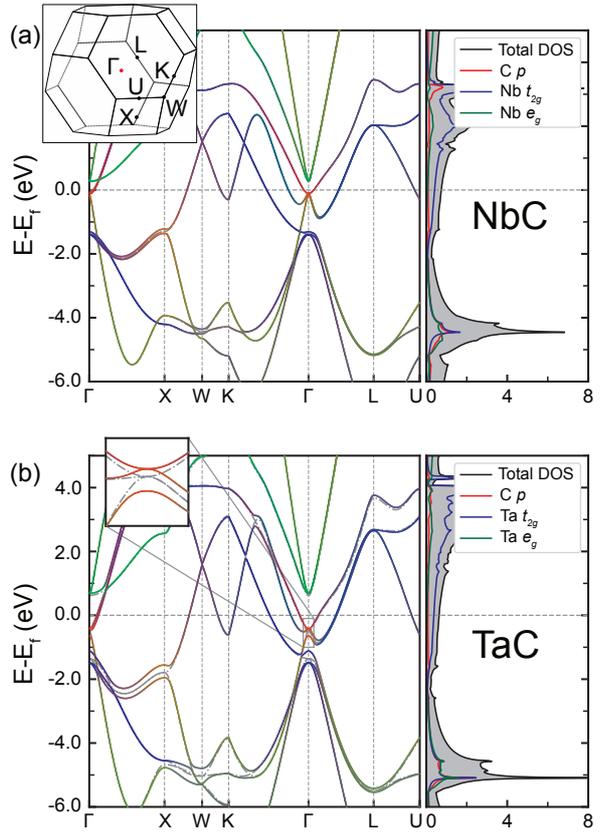}
	\caption{Electronic band structures of (a) NbC and (b) TaC, 
		calculated by considering (solid colored lines) and by 
		ignoring (dash-dotted grey lines) the spin-orbit coupling. 
		The $d$-orbitals in Nd or Ta and the $p$-orbitals in C are 
		presented in blue and red colors, respectively. The inset in (b) 
		shows the detailed band structure around the $\Gamma$ point. 
		The total- and partial (Nb or Ta and C atoms) density of states 
		with SOC are shown on the right side of the panel.
		The primitive cell Brillouin zone, including the high-symmetry 
		points, is shown on the top panel.}
	\label{fig:DOS}
\end{figure}
%

\subsection{\label{ssec:Cp_zero} Zero-field specific heat}

The zero-field electronic specific-heat data 
were further analyzed.  
As shown by the dashed lines in the insets of Fig.~\ref{fig:Cp}, 
the normal-state specific heat was fitted to $C/T = \gamma_\mathrm{n} + \beta T^2 + \delta T^4$, 
where $\gamma_\mathrm{n}$ is the normal-state electronic specific heat coefficient, 
and the two other terms account for the phonon contribution to the specific heat. 
After subtracting the phonon contribution from the experimental data, the electronic specific heat divided by the electronic specific-heat 
coefficient, i.e., $C_\mathrm{e} / \gamma_\mathrm{n} T$, is shown in the main panels of Fig.~\ref{fig:Cp} vs.\ the reduced temperature $T/T_c$. 
The solid lines in Fig.~\ref{fig:Cp} represent fits with $\gamma_\mathrm{n}$ =  3.0(4) and 3.1(8)\,mJ/mol-K$^2$ and a single isotropic gap 
$\Delta_0 = 1.73(2)$ and 1.54(2)\,meV for NbC and TaC, respectively. 
They reproduce very well the experimental data, while being comparable 
with the TF-$\mu$SR results (see Fig.~\ref{fig:lambda}).

The Debye temperature $\Theta_\mathrm{D}$ can be calculated 
by using $\Theta_\mathrm{D} = (12\pi^4\,Rn/5\beta)^{1/3}$, 
where $R = 8.314$\,J/mol-K is the molar gas constant and $n = 2$ is the number of atoms per formula unit. 
%
%
\begin{figure}[ht]
	\centering
	\includegraphics[width = 0.44\textwidth]{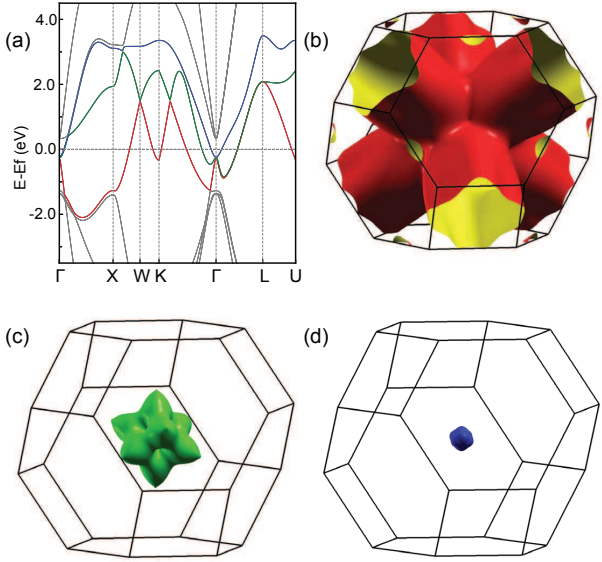}
	\caption{(a) Close up view of the NbC electronic band structure. 
		The bands crossing the Fermi level are highlighted in red, green, 
		and blue. (b)-(d) Representative Fermi surfaces of NbC using the 
		same color code of the bands shown in (a). Very similar Fermi 
		surfaces were found also in the TaC case.}
	\label{fig:FS_NbC}
\end{figure}
%
%
With $\beta$ = 8.0(8) and 19(2)\,$\mu$J/mol-K$^4$, the estimated  
$\Theta_\mathrm{D}$ are 790(20) and 590(20)\,K for NbC and TaC, respectively. 
The density of states (DOS) at the Fermi level $N(\epsilon_\mathrm{F})$ was evaluated 
from the expression $N(\epsilon_\mathrm{F}) = 3\gamma_\mathrm{n}/(\pi^2 k_\mathrm{B}^2)$ $\sim$ 1.3\,states/eV-f.u. for both NbC and TaC~\cite{Kittel2005}, where 
$k_\mathrm{B}$ is the Boltzmann constant. 
The electron-phonon coupling constant $\lambda_\mathrm{ep}$ 
was estimated from the $\Theta_\mathrm{D}$ and $T_c$ values by using the 
semi-empirical  McMillan formula~\cite{McMillan1968}:
\begin{equation}
\lambda_\mathrm{ep}=\frac{1.04+\mu^{\star}\,\mathrm{ln}(\Theta_\mathrm{D}/1.45\,T_c)}{(1-0.62\,\mu^{\star})\mathrm{ln}(\Theta_\mathrm{D}/1.45\,T_c)-1.04}.
\end{equation}
The Coulomb pseu\-do\-po\-ten\-tial $\mu^{\star}$, typically lying in the 0.09--0.18 
range, was fixed here to 0.13, a commonly used value for 
transition metals. From the above expression, we obtain $\lambda_\mathrm{ep}$ = 0.60(1) and 0.65(3) for NbC and TaC, 
both consistent with theoretical values~\cite{Isaev2007}. 
Finally, the band-structure density of 
states $N_\mathrm{band}(\epsilon_\mathrm{F})$ can be estimated from the relation 
$N_\mathrm{band}(\epsilon_\mathrm{F}) = N(\epsilon_\mathrm{F})/(1 + \lambda_\mathrm{ep}$)~\cite{Kittel2005}, 
which gives $N_\mathrm{band}(\epsilon_\mathrm{F})$ $\sim$ 0.8\,states/eV-f.u.\ for both compounds. 

\subsection{\label{ssec:DFT}Electronic band-structure calculations}

To further understand the electronic properties of NbC and TaC, we also performed DFT calculations.
The electronic band structure results, as well as the density of states (DOS) are depicted in Fig.~\ref{fig:DOS}. 
Since NbC and TaC adopt
the same rock-salt structure (see Fig.~\ref{fig:XRD}), 
it is natural that their band structures and DOS profiles are quite similar.
Close to the Fermi level the bands are dominated by the $d$-electrons of TMs (Nb or Ta), while the contribution from the C $p$-electrons is quite modest. The $d$- and $p$-electrons are highly hybridized. The bands which cross the Fermi level are mainly occupied by the $t_{2g}$ ($d_{xy}$, $d_{yz}$, and $d_{xz}$) orbitals of TM,  with marginal contributions from the $e_g$- ($d_{x^2-y^2}$ and $d_{z^2}$) 
or the $p$-orbitals of C. The occupied $e_g$ bands are mainly located about 4.0 eV below the Fermi level, while the unoccupied bands extend
from 0.5\,eV to above 5.0\,eV. 
In TaC, the sixfold degenerate point at the $\Gamma$ point below 
the Fermi level is split into one doubly-degenerate and one four-degenerate 
point by the spin-orbit coupling, as shown in the inset of Fig.~\ref{fig:DOS}(b).
In the lighter NbC, the SOC splitting is less evident. 
Also the bands along the $\Gamma$-$X$ direction are split, the maximum 
band splitting $E_\mathrm{SOC}$ being about 400\,meV in TaC, which is 
much larger than 130\,meV in NbC. In both cases, the band splitting is 
caused by the dominant SOC of the Ta or Nb $d$-electrons.
As shown in the right panels of Fig.~\ref{fig:DOS}, the estimated DOS 
at the Fermi level is $\sim$0.74\,states/eV-f.u.\ for NbC, slightly 
higher than 0.64\,states/eV-f.u.\ in TaC. Both values are comparable to 
the experimental value calculated from the electronic specific-heat 
coefficient (see Sec.~\ref{ssec:Cp_zero} and Table~\ref{tab:parameter}).

\begin{figure}[!htp]
	\centering
	\includegraphics[width = 0.46\textwidth]{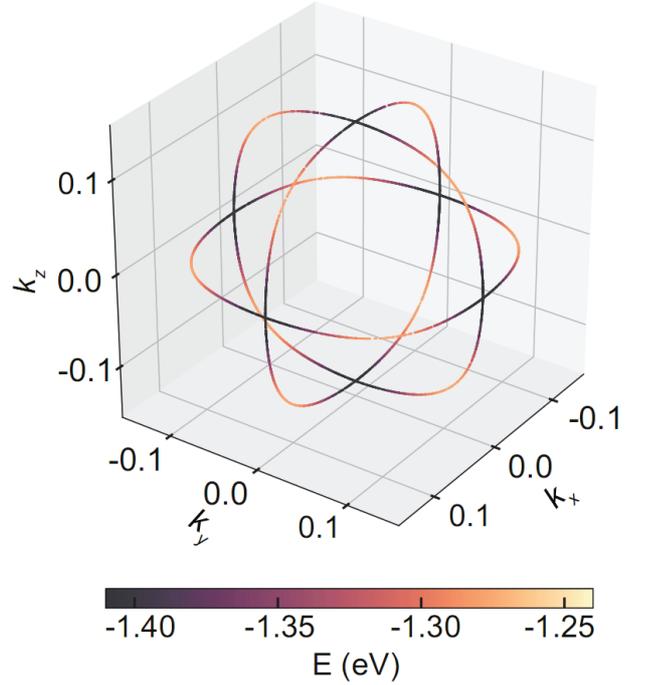}
	\caption{Nodal loops around the $\Gamma$ point in NbC, with TaC 
		showing similar features. The color coding reflects the energy 
		scale, indicated by the colorbar at the  bottom.}
	\label{fig:Nodal}
\end{figure}

There are three bands crossing the Fermi level, highlighted by 
different colors in Fig.~\ref{fig:FS_NbC}(a) for the NbC case.  
We depict each Fermi surface (FS) of these three bands in 
Fig.~\ref{fig:FS_NbC}(b)-(d), respectively. Since TaC exhibits a 
similar band structure, here we show only the Fermi-surface plots of NbC.
The three bands form three distinct electron FSs, one of which 
consists of three large cylinders along the $k_x$, $k_y$, and $k_z$ directions 
[see Fig.~\ref{fig:FS_NbC}(b)]. Such large cylinder-like FSs originate from the hybridization between the TM $t_{2g}$-orbitals and C $p$-orbitals. The cylinder-like FSs are known to play an important role in the superconductivity of high-$T_c$ iron-based materials~\cite{Mazin2008,Kuroki2009,Mazin2019}. 
This might be also the case of NbC and TaC carbides, both of which exhibit relatively high superconducting temperatures.
The other two small FSs shown in Figs.~\ref{fig:FS_NbC}(c) and (d) are more 
three-dimensional-like compared to the large FS in Fig.~\ref{fig:FS_NbC}(b).

After checking the band structure (without SOC) across the whole 
Brillouin zone, we could identify three closed nodal loops, lying in 
the $k_x = 0$, $k_y = 0$, and $k_z = 0$ planes and centered at 
the $\Gamma$ point. 
These are shown in Fig.~\ref{fig:Nodal} for the NbC case, 
with TaC displaying similar features. 
We recall that NbC and TaC have the same rock-salt structure, 
with a space group $Fm\overline{3}m$ (No.\,225), characterized by three mirror planes, i.e., $m_{xy}$, $m_{yz}$, and $m_{xz}$.
According to a symmetry analysis~\cite{Zeng2015}, the nodal loops are protected by the three mirror symmetries, since the crossing bands belong to different mirror eigen\-values.
When considering SOC, the nodal loops become gapped, except for the six Dirac points $(\pm k_x,0,0)$, $(0,\pm k_y, 0)$, and $(0,0,\pm k_z)$, which are protected by the $C_4$ and the combined 
space-time inversion $PT$ symmetry. \tcr{Clearly, this is consistent with 
	the preserved time-reversal symmetry we find from the ZF-$\mu$SR results}.
Since Nb atoms exhibit a weaker intrinsic SOC, we propose that NbC should 
be a good candidate for studying the exotic 2D surface states, as well 
as topological superconductivity~\cite{Zhao2016}.

\section{Discussion}
\tcr{The possibilities offered by topological superconductors, ranging from 
	hosting Majorana fermion quasiparticles to potential applications in 
	topological quantum computing~\cite{Qi2011,Kitaev2001,Liang2011}, have 
	spurred the researchers to explore different routes in the search for 
	materials that can realize them. 
	One approach consists in combining conventional $s$-wave superconductors 
	with topological insulators to form heterostructures. The resulting 
	proximity effect between the respective surface states 
	can lead to a two-dimensional superconducting state with 
	$p+ip$ pairing, known to support Majorana bound states at 
	the vortices~\cite{Liang2008}. 
	For instance, evidence of topological SC has been 
	reported in NbSe$_2$/Bi$_2$(Se,Te)$_3$~\cite{Xu2014,Peng2014}, where 
	NbSe$_2$ represents a typical fully-gapped superconductor, while 
	Bi$_2$(Se,Te)$_3$ are both well known topological insulators.
	Regrettably, the complexity and difficulty of fabricating such 
	heterostructures and their relatively low superconducting transition 
	temperatures (typically below 4\,K) limit further studies and possible 
	applications. 
	To achieve topological superconductivity, one can also consider introducing 
	extra carries into a topological insulator, as e.g., in Cu intercalated 
	Bi$_2$Se$_3$~\cite{Hor2010,Sasaki2011}. Again, doping-induced 
	inhomogeneities and disorder effects hinder further investigations of 
	these doped topological insulators.
}

\tcr{A more attractive 
	route to attain topological superconductivity is by 
	combining SC with a nontrivial electronic band structure 
	into the same compound. Clearly, it is of fundamental interest to be able 
	to identify such new superconductors with nontrivial band topology, 
	yet with a simple composition and high transition temperatures. For example, 
	topologically protected surface states have been found in superconducting 
	$\beta$-PdBi$_2$ and PbTaSe$_2$~\cite{Guan2016,Sakano2015}, both representing 
	good platforms for studying topological SC. Unfortunately, 
	their transition temperatures are still relatively low ($< 5$\,K).}

\tcr{In this study, we found that both NaC and TaC could be potentially 
	interesting materials, where a nontrivial topological band structure 
	coexists with superconductivity. By using both macroscopic and microscopic 
	techniques, we found that the rock-salt-type NbC and TaC exhibit relatively 
	high superconducting transitions (at $T_c$ = 11.5 and 10.3\,K). 
	The low-temperature superfluid density and electronic specific-heat both 
	suggest a fully-gapped superconducting state in NbC and TaC. 
	The numerical band-structure calculations indicate that without considering 
	the SOC effect, the first Brillouin zone contains three closed node lines 
	in the bulk band structure, protected by time-reversal and space-inversion 
	symmetry, which is consistent with the ZF-$\mu$SR results (see Fig.~\ref{fig:ZF_muSR}). 
	When considering SOC, this picture might change. However, since SOC 
	effects in the NbC case are rather weak, it is highly probable that the 
	node lines are still preserved. Should this be confirmed by future 
	investigations, considering also its relatively high $T_c$, NbC would be a 
	very interesting topological superconductor.}

%
%

%
\begin{table}[!th]
	\centering
	\caption{Normal- and superconducting state properties of NbC and TaC, as 
		determined from magnetization, specific-heat, and $\mu$SR measurements, 
		as well as electronic band-structure calculations.\label{tab:parameter}}
	\begin{ruledtabular}
		\begin{tabular}{lccc}
			Property                               & Unit            & NbC        & TaC  \\ \hline
			$T_c^\chi$                             & K               & 11.5       & 10.3   \\
			$T_c^C$                                & K               & 10.6       & 10.2   \\
			$\mu_0H_{c1}^\chi$                          & mT              & 10.3(3)    & 29.3(3)  \\
			$\mu_0H_{c2}$$^{\chi,C}$               & T               & 1.93(1)   & 0.65(1) \\
			$\mu_0H_{c2}$$^{\mu\mathrm{SR}}$\footnotemark[1]         & T         & 1.81(2)   & 0.43(3)   \\
			$\xi(0)^{\chi,C}$                               & nm              & 13.1(1)   & 22.5(1)  \\[2mm]
			$\gamma_n^C$                             & mJ/mol-K$^2$    & 3.0(4)    & 3.1(8)  \\
			$\Theta_\mathrm{D}^C$                    & K               & 790(20)   & 590(20)   \\
			$\lambda_\mathrm{ep}^C$                  & ---             & $\sim$0.60(1)   & $\sim$0.65(3)  \\
			$N(\epsilon_\mathrm{F})^C$               & states/eV-f.u.  & 1.3(2)    & 1.3(3)    \\
			$N_\mathrm{band}(\epsilon_\mathrm{F})^C$ & states/eV-f.u.  & 0.8(1)    & 0.8(1)  \\
			$N(\epsilon_\mathrm{F})^\mathrm{DFT}$  & states/eV-f.u.  & 0.74      & 0.64  \\  
			$E_\mathrm{SOC}^\mathrm{DFT}$                       & meV             & 130        & 400       \\[2mm]
			$\Delta_0$$^{\mu\mathrm{SR}}$          & meV             & 1.90(2)    & 1.45(1)  \\  
			$\Delta_0^{C}$                         & meV             & 1.73(2)    & 1.54(2)  \\
			$\lambda_0^{\mu\mathrm{SR}}$\footnotemark[1]            & nm              & 134(2)     & 67(1)   \\
			$\lambda_0^{\mu\mathrm{SR}}$                            & nm              & 141(2)     & 77(1)   \\
		\end{tabular}
		\footnotetext[1]{Derived from a fit to Eq.~\eqref{eq:TF_muSR_H} at 1.5\,K.}	
	\end{ruledtabular}
\end{table}
%

\section{\label{ssec:Sum}Conclusion}
To summarize, we studied the superconducting properties of the 
NbC and TaC superconductors by means of bulk- (magnetization 
and heat capacity) and local-probe ($\mu$SR) techniques, as well as via numerical 
band-structure calculations. The superconducting state of NbC and TaC is 
characterized by $T_c$ = 11.5\,K and 10.3\,K, and upper critical fields 
$\mu_0H_{c2}$ = 1.93\,T and 0.65\,T, respectively. 
The temperature dependence of the superfluid density 
and the zero-field electronic specific heat reveal a \emph{nodeless} 
superconducting state, which is well described by an 
\emph{isotropic $s$-wave} model. 
The lack of spontaneous magnetic fields below $T_c$ indicates 
that time-reversal symmetry is \emph{preserved} in the superconducting state of 
NbC and TaC. Electronic band-structure calculations suggest that the 
density of states at the Fermi level stems primarily
from the Nb (or Ta) $d$-electrons and the C $p$-electrons. 
The strong hybridization between the C $p$-orbitals and Nb (or Ta) $t_{2g}$ orbitals produces large cylinder-like Fermi surfaces, 
resembling those 
of high-$T_c$ iron-based superconductors. 
Three closed node lines are found in the first Brillouin zone, which are protected by time-reversal and inversion symmetry in the band structure 
of the bulk. 
In particular, we show that NbC 
is a potential candidate for future studies of  
topological superconductivity.

\begin{acknowledgments} 
	This work was supported by the Schwei\-ze\-rische Na\-ti\-o\-nal\-fonds 
	zur F\"{o}r\-de\-rung der Wis\-sen\-schaft\-lich\-en For\-schung, SNF 
	(Grants no.\ 200021-169455 and 206021-139082). The $\mu$SR experiments 
	were performed at the $\pi$M3 beamline of the Paul Scherrer Institute. 
	We thank the scientists of the GPS $\mu$SR spectrometer for their support.
\end{acknowledgments}

\bibliography{TmC_bib}

\end{document}